\documentclass{article}
\usepackage{spconf,amsmath,epsfig}
\usepackage{graphicx}
\usepackage{subcaption}
\usepackage{booktabs}
\usepackage{multirow}                 
\usepackage{multicol}
\usepackage{cite}
\usepackage{bigstrut}
\usepackage[table]{xcolor}
\usepackage[colorlinks = true, linkcolor=red,
    filecolor=blue,
    urlcolor=black,
    citecolor=green,
    ]{hyperref}
\usepackage{bbding}
\usepackage{pifont}
\usepackage{dsfont}
\usepackage{amsmath,bm}
\usepackage{amssymb,stmaryrd}

\newcommand{\cmark}{\ding{51}}%
\newcommand{\xmark}{\ding{55}}%

\usepackage[misc]{ifsym}
\let\OLDthebibliography\thebibliography
\renewcommand\thebibliography[1]{
  \OLDthebibliography{#1}
  \setlength{\parskip}{0pt}
  \setlength{\itemsep}{0pt plus 0.3ex}
}

\begin{document}\sloppy

\def\x{{\mathbf x}}
\def\L{{\cal L}}

\title{Enhancing Few-Shot Classification without Forgetting through Multi-Level Contrastive Constraints
}
%
\name{Bingzhi Chen$^{\dagger  }$ \quad Haoming Zhou$^{\dagger}$ \thanks{*B. Chen and H. Zhou are co-first authors of the article.} \quad Yishu Liu$^{\ddagger} $  \quad Biqing Zeng$^{\dagger}$ \quad Jiahui Pan$^{\dagger \, \scalebox{0.5}{\Letter}} \thanks{\Letter  \, Corresponding author.}$ \quad Guangming Lu$^{\ddagger}$  }
\address{$^{\dagger}$ South China Normal University, Guangzhou, China \\
$^{\ddagger}$ Harbin Institute of Technology, Shenzhen, China
}

\maketitle

\begin{abstract}
Most recent few-shot learning approaches are based on meta-learning with episodic training. However, prior studies encounter two crucial problems: (1) \textit{the presence of inductive bias}, and (2) \textit{the occurrence of catastrophic forgetting}. In this paper, we propose a novel Multi-Level Contrastive Constraints (MLCC) framework, that jointly integrates within-episode learning and across-episode learning into a unified interactive learning paradigm to solve these issues. Specifically, we employ a space-aware interaction modeling scheme to explore the correct inductive paradigms for each class between within-episode similarity/dis-similarity distributions. Additionally, with the aim of better utilizing former prior knowledge, a cross-stage distribution adaption strategy is designed to align the across-episode distributions from different time stages, thus reducing the semantic gap between existing and past prediction distribution. Extensive experiments on multiple few-shot datasets demonstrate the consistent superiority of MLCC approach over the existing state-of-the-art baselines.
\end{abstract}
\section{Introduction}
\label{sec:intro}

Deep learning has achieved tremendous success in the field of computer vision and has been applied in numerous practical scenarios, primarily due to its access to extensive training data. However, in many practical cases, i.e., cold-start recommendation~\cite{zheng2021cold} and medical diagnose~\cite{prabhu2019few}, only a small number of training samples are available, posing a challenge of achieving relatively ideal performance with limited sample support. In recent years, numerous scholars~\cite{sung2018learning, vinyals2016matching, snell2017prototypical} have been dedicated to researching few-shot learning, achieving significant breakthroughs in the field.

Performing few-shot classification well is essential towards creating robust frameworks that can learn with the efficiency of humans. In many cases, an effective and viable research line involves both meta-learning and metric learning. In this approach, meta-learning is formulated to employ episodic training techniques to partition a large base-class dataset into numerous episodes. By learning the meta-knowledge of each episode as prior knowledge, it aims to enhance generalization to novel-class samples. To acquire highly generalizable feature representations through the neural networks, the metric-based methods employ a similarity function between query images and support samples. In few-shot classification tasks, the fusion of episodic training techniques and metric learning methods has been demonstrated to achieve remarkable performance~\cite{afrasiyabi2022matching,xie2022joint,zhang2022metanode}.

\begin{figure}[t]
  \centering
  \begin{subfigure}{1\linewidth}
     \includegraphics[width=1\linewidth]{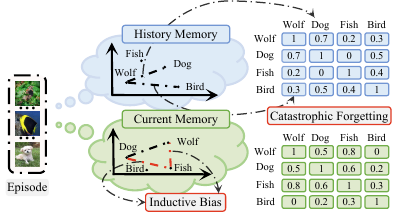}
  \end{subfigure}
  \caption{Illustration of the inductive bias and the catastrophic forgetting during meta-training episode.}
  \vspace{-0.2in}
  \label{Fig_1}
\end{figure}

    
    
  


However, most existing methods based on meta-learning and metric learning encounter two crucial problems, i.e., the {\bf inductive bias}~\cite{liu2021learning} and the {\bf catastrophic forgetting}~\cite{kemker2018measuring}, which are illustrated in Fig.~\ref{Fig_1}. On the one hand, during episode training, due to limited training samples, the learner confuses more closely related categories, e.g., if images of dogs and wolves are highly correlated, a learning model trained on such data might tend to associate images resembling dogs with wolves. With only a few samples, the model can easily learn inductive bias, resulting in overfitting. On the other hand, during continual episode learning, catastrophic forgetting manifests as the learner gradually relies on new episode information, subsequently forgetting the information from previous episodes. Catastrophic forgetting drives the few-shot model further diminish its generalization to unseen-category or novel-class samples. 

To address the issues mentioned above, we propose a novel Multi-Level Contrastive Constraints (MLCC) framework, which jointly integrates within-episode learning and across-episode learning into a unified interactive learning paradigm for enhancing few-shot classification against inductive bias and catastrophic forgetting. With the aim of removing the error accumulation caused by inductive bias, we propose a space-aware interaction modeling scheme to guide the model in learning the correct inductive paradigm for each class by utilizing within-episode similarity/dis-similarity distributions. This ensures that the semantic distribution extracted from images through deep networks is dynamically realigned towards class-specific information, leading to more distinctive embedding representations. Furthermore, to alleviate the problem of catastrophic forgetting, a cross-stage distribution adaption strategy is designed to align the across-episode distributions from different time stages within the paradigm of continual meta-learning. This strategy is expected to effectively reduce the semantic gap between existing and past prediction distribution. As such, 
we conduct sufficient experiments to validate the competence of our method on multiple few-shot learning datasets. The superb recognition performance against state-of-the-art methods on multiple visual classification tasks, e.g., general object categorization and fine-grained recognition, demonstrates the robustness and efficacy of our approach in addressing few-shot classification. 
Overall, our main contributions are summarized as follows:
\begin{itemize}
  \item  We propose a novel few-shot classification framework called MLCC, which integrates within-episode learning and across-episode learning into a unified interactive learning paradigm, offering a promising solution against inductive bias and catastrophic forgetting.
  \item To remove the error accumulation caused by inductive bias, we propose a space-aware learning modeling scheme that utilizes within-episode similarity/dis-similarity distributions to learn the correct inductive paradigms for each class within the meta-learning.
  \item To alleviate the catastrophic forgetting, we designed a cross-stage distribution adaption strategy to align the across-episode distributions from different time stages, which can effectively reduce the semantic gap between existing and past prediction distribution.
  \item We evaluate our MLCC framework on multiple few-shot learning datasets. The superb performance collectively demonstrates the effectiveness and superiority of our method for addressing few-shot classification.
\end{itemize}

%
%

\section{Related Works}
\subsection{Few-shot Learning}
Few-shot learning aims to recognize novel classes with few labeled samples by adapting the prior knowledge learned from the base dataset. 
Much effort has been devoted by recent work, which can be roughly grouped into two categories. 1) Metric-based approaches. This line of works aims at improving how the distance is calculated for better performance at training and inference. ProtoNet~\cite{snell2017prototypical} learned to classify samples by comparing their euclidean distance to the prototypes.  2) Optimization-based approaches. The key idea is to acquire knowledge on optimizing the gradient descent process, ensuring effective initialization or update direction for the meta-learner. MAML~\cite{finn2017model} aimed to learn a good parameter initialization to facilitate faster adaptation to novel tasks. 

\subsection{Catastrophic Forgetting}
Due to the stability-plasticity dilemma, catastrophic forgetting~\cite{kemker2018measuring} has hindered further development in deep neural networks. As this issue gradually receives attention, recently proposed methods can be roughly cast into two categories. 1) Regularization approaches~\cite{li2017learning} use various constraints to reduce memory forgetting. 2) Dual-memory algorithms~\cite{soltoggio2018born} store and reuse a small amount of raw data in a memory buffer. Being more related to the second category, we propose a cross-stage distribution adaption strategy to strengthen the memory of past prior knowledge.

\subsection{Contrastive Learning}
Over the past few years, contrastive learning has been widely used in few-shot classification, which aims to learn an embedding space by maximizing the similarities of positive pairs and minimizing those of the negatives. Inspired by the idea of contrastive learning, InfoPatch~\cite{liu2021learning} applied supervised contrastive loss in both the pre-training and meta-training stages to learn more effective embeddings. SCL~\cite{ouali2021spatial} presented a novel attention-based spatial contrastive objective to learn locally discriminative and class-agnostic features. Unlike prior works, our proposed method leverages multi-level contrastive training scheme to boost few-shot classification performance solely within the meta-training stage.


\section{Proposed Approach}
\label{sec:Problem Definition}

\begin{figure*}[t]
  \centering
  %
    \includegraphics[width=1\linewidth]{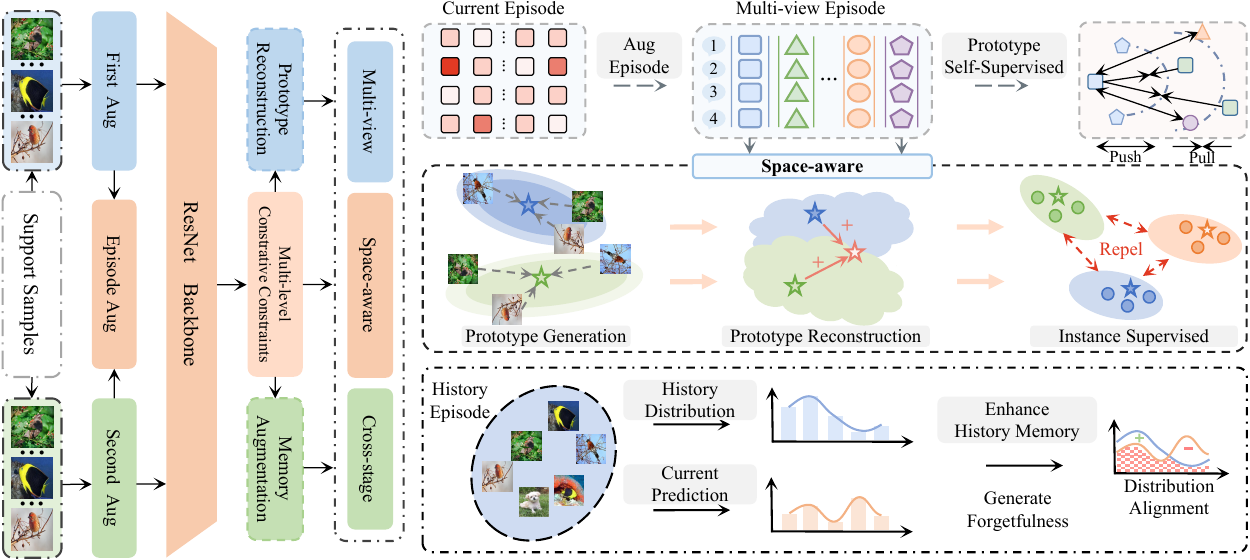}
    \caption{Illustration of the proposed Multi-Level Contrastive Constraints (MLCC) framework for boosting few-shot classification.}
    \label{fig:main_fig}
  \hfill
  \vspace{-0.2in}
\end{figure*}
\subsection{Problem Definition}

We follow the standard protocol to define few-shot classification paradigm. Given the based dataset $\mathcal{D}_\text{base}$, our goal is to learn a model through episodic meta-training strategy that can accurately recognize the novel dataset $\mathcal{D}_\text{novel}$ with only a few labeled examples in each class, where $\mathcal{D}_\text{base} \cap \mathcal{D}_\text{novel}=\varnothing$. Technically, episodic meta-training can be formulated as a range of $N$-way $K$-shot episode problems, e.g., an episode consists of a support set $\mathcal{D}_\text{sup}={(x_i, y_i)}_{i=0}^{N \times K}$ and a query set $\mathcal{D}_\text{que}={(x_i, y_i)}_{i=0}^{N \times Q}$, where $N$ denotes the number of classes, $K$ and $Q$ represent the amount of images in the support set and query set for each class, respectively. 



\subsection{Overview of MLCC}
The proposed multi-level contrastive constraints framework integrates within-episode and across-episode learning into a unified interactive learning paradigm, which can allow the model to explore more distinctive and shared semantic information between two types of episodic learning. Fig.~\ref{fig:main_fig} provides a detailed pipeline of the proposed MLCC framework, comprising three essential modules:  1) Multi-view Episode Scheme, 2) Space-aware Interaction Modeling, and 3) Cross-stage Distribution Adaption. Through the joint optimization of these modules, our approach can
further improve the performance and reliability of few-shot models.


\subsection{Multi-view Episode Scheme}
To implicitly remove the inductive bias phenomenon, our approach aims to capture diverse patterns and variations from multi-view episodic augmentations, which can introduce diversity and robustness to within-episode feature representations for each support class within the meta-training stage.

\textbf{Multi-view Episodic Augmentation:} Following the idea of contrastive learning~\cite{yang2022few}, our MLCC framework incorporates an augmentation technique, i.e., constructing multi-view episode by two separate data augmentation operators from the family of augmentations ($t\sim\mathcal{T}$ and $t^{\prime}\sim\mathcal{T}$). The principle of constructing multi-view episode can be formulated as:
\begin{equation}
\small
\begin{aligned}
E_1= t(E), E_2= t^{\prime}(E),
\end{aligned}
\end{equation}
where $E$ is the episode randomly sampled from meta-training set, and $E_i$ represents the $i$-th episode view from $E$.

\textbf{Multi-view Episodic Prototype:}
For the purpose of digging out inner contextual shared information and preserving pairwise semantic similarities from multi-view episodic feature embeddings, we designed multiple episodic prototypes from two separate augmentation episodes and a mixup-style hybrid prototype for each support class during the episodic training process, which aims to assist in space-aware interaction modeling, i.e., prototype generation, prototype reconstruction, and inter/intra-class contrastive constraints. Therefore, the multiple episodic prototypes and mix-up style hybrid prototype can be respectively computed by:
\begin{equation}
\small
\mathcal{P}_n^v=\frac1K\sum_{i=1}^K\mathcal{F}(x_{n,i}^v),
\end{equation}
\begin{equation}
\small
\widetilde{\mathcal{P}}_n=\alpha\cdot\mathcal{P}_n^1+(1-\alpha)\cdot\mathcal{P}_n^2,
\label{Eq_3}
\end{equation}
where $v\in\{1,2\}$, $\mathcal{P}^{v}_{n}$ denotes the class prototype belongs to class $n$ from the support set under the $v$-th episode view, $\widetilde{\mathcal{P}}_n$ denotes the mix-up style hybrid prototype, and $\alpha$ denotes a tunable weight parameter that controls the mixing strength.

\subsection{Space-aware Interaction Modeling}
During meta-learning, excellent category inductive ability is necessary for extremely limited training examples. Our proposed space-aware interaction modeling scheme is designed to explore the similarity/dis-similarity relationship between the multi-view episodic prototypes and instances in feature distributions to alleviate the induction bias.


\textbf{Inter-class Space Modeling:} We introduced an inter-class contrastive constraint that combines multiple episodic prototypes from multi-view episodic prototypes to encourage the network to generate more distinctive class prototypes from feature embeddings of each semantic class. This constraint aims to capture discriminative insights and improve inductive power of feature clusters across various semantic categories. The inter-class loss can be defined as follows:
{\small
\begin{equation}
\mathcal{L}_{\text{Inter-class}}=-\sum_{i\in N}\log\frac{\exp(\cos(\mathcal{P}^{1}_i,\mathcal{P}^{2}_i)/\kappa)}
{\sum_{v} \sum_{j\in N} \mathds{1}_{j\neq i} \exp(\cos(\mathcal{P}^{1}_i,\mathcal{P}^{v}_j)/\kappa)},
\end{equation}
}
where $\kappa$ denotes a tunable temperature parameter. Following that, we can clearly distinguish differences and realize class-specific details between inter-class clusters.

\textbf{Intra-class Space Modeling:} 
To explore more detailed and precise clusters for each specific class, an intra-class contrastive loss is incorporated to bring instances belonging to the same class closer in the latent embedding space. In particular, it combines a mix-up style prototype from multi-view episodic prototypes, aiming to enhance the inductive ability of the learned representations for each semantic category. Formally, the intra-class loss is defined as follows:
\begin{equation}
\small
\mathcal{L}_{\text{Intra-class}}=-\log\frac{\exp(\cos(\mathcal{F}(x),\widetilde{\mathcal{P}}_{y})/\tau)}{\sum_{n=1}^N\exp(\cos(\mathcal{F}(x),\widetilde{\mathcal{P}}_n)/\tau)},
\end{equation}
where $\tau$ denotes a scalable temperature parameter. Later on, the samples of each class can be made similar around semantic concepts by intra-class inductive learning paradigm. 
\subsection{Cross-stage Distribution Adaption}
From the perspective of mitigating the catastrophic forgetting, a cross-stage distribution adaption strategy is meticulously built on the principles of maximal margin clustering~\cite{huang2020deep}, to promote the model's current prediction distribution to align with the historical distribution for same episode across different time stages within the paradigm of meta-learning. 

\textbf{Across-episode Priors Learning:} 
To enhance the memory capacity of past priors in feature clusters across different semantic classes, we strategically introduce an against-forgetting loss that is designed to capture both local and global information between past and present prior knowledge among semantic categories. Finally, the against-forgetting loss can be mathematically computed as follows:
\begin{equation}
\small
\begin{aligned}
\mathcal{L}_\text{Forget} = & \frac{1}{N}\sum_{n=1}^{N} \Bigg[ -\log\left(\frac{\mathcal{A}^n}{\|\mathcal{A}\|_2}\cdot\frac{\mathcal{H}^n}{\|\mathcal{H}\|_2}\right) \\
& -\left(\frac{\sum\mathcal{A}^n}{\sum\mathcal{A}}*\log(\frac{\sum\mathcal{A}^n}{\sum\mathcal{A}})\right)\Bigg] + \delta , \\
& \mathrm{s.t.} \quad \begin{aligned}[t] 
&\mathcal{A}=[\mathcal{A}_1,\mathcal{A}_2,...,\mathcal{A}_{N}], \\
&\mathcal{H}=[\mathcal{H}_1,\mathcal{H}_2,...,\mathcal{H}_{N}],
\end{aligned}
\end{aligned}
\end{equation}
where $\mathcal{A}\in\mathbb{R}^{N\times Q}$ and $\mathcal{H}\in\mathbb{R}^{N\times Q}$ respectively represent the current reidentification distribution and previous prediction distribution for same episode from historical episodic cache information, $\delta$ denotes an additional regularization term that is used to prevent negative loss values during training process.

\subsection{Objective Optimization}
Based on the end-to-end training scheme, the total learning objective for the proposed MLCC approach is formulated as:
\begin{equation}
\small
\mathcal{L}_{\text{Total}}=\underbrace{\mathcal{L}_{\text{CE}}}_{\text{Classification}}+\underbrace{(\lambda_1\cdot{\mathcal{L}_{\text{Inter-class}}}+\lambda_2\cdot{\mathcal{L}_{\text{Intra-class}}})}_{\text{Space-aware}}+\underbrace{\mathcal{L}_{\text{Forget}}}_{\text{Cross-stage}},
\end{equation}
where $\lambda_1=2$ and $\lambda_2=1$ denote the importance of the inter/intra-class loss, and $\mathcal{L}_{\text{CE}}$ is cross-entropy (CE) loss.

\section{Experiment}
\label{sec:experiment}

\subsection{Datasets and Metrics}
Following the purpose of comprehensively evaluating the effectiveness and generalizability of the proposed algorithm, we performed few-shot image classification tasks on a variety of image benchmark datasets, including the general object recognition benchmarks, i.e., miniImageNet \cite{vinyals2016matching} and tieredImageNet \cite{ren2018meta}, and fine-grained categorization datasets, i.e., CUB \cite{wah2011caltech}. Additionally, we extend experiments on domain transfer conditions introduced by judging performance on Cars \cite{krause20133d}, Aircraft \cite{maji2013fine}, and CUB after training on miniImageNet. Same as previous works \cite{xie2022joint, dong2023exploring}, we report the average accuracy (\%) and the corresponding 95\% confidence intervals over 2000 test episodes after applying our method on 5-way 1-shot and 5-way 5-shot settings for mentioned datasets.


\begin{table*}[t]
  \small
  \centering
  \renewcommand\tabcolsep{2.8pt}
  \renewcommand\arraystretch{1.1}
  \caption{Comparison with the state-of-the-art methods for both general and fine-grained few-shot image classification.}
  \vspace{-0.1in}
  \label{Tab_1}
  \begin{tabular}{l|c|cc|cc!{\vline width 0.8pt}l|c|cc}
  \toprule
    
    \multicolumn{1}{l|}{\multirow{2}{*}{Methods}} 
    & \multicolumn{1}{c|}{\multirow{2}{*}{ResNet-12}} 
    & \multicolumn{2}{c|}{miniImageNet} 
    & \multicolumn{2}{c!{\vline width 0.9pt}}{tieredImageNet}
    &\multicolumn{1}{l|}{\multirow{2}{*}{Methods}} 
    & \multicolumn{1}{c|}{\multirow{2}{*}{ResNet-12/18}} 
    & \multicolumn{2}{c}{CUB}\\
    \multicolumn{1}{l|}{}  &  &5way-1shot  & 5way-5shot  & 5way-1shot & 5way-5shot &\multicolumn{1}{l|}{}  &  & 5way-1shot & 5way-5shot \bigstrut[b]\\
    
    \hline
    ProtoNet \cite{snell2017prototypical}  &  NIPS’17   & 62.11±0.44&80.77±0.30 &68.31±0.51  
    &83.85±0.36  & GEmbed* \cite{tian2020rethinking} &  ECCV’20   & 77.92±0.46 & 89.94±0.26  \bigstrut[t]\\



    HGNN \cite{yu2022hybrid} &  AAAI’22   & 67.02±0.20 & 83.00±0.13 & 72.05±0.23 & 86.49±0.15  & ADM* \cite{li2020asymmetric} &   IJCAI’20  & 79.31±0.43 & 90.69±0.21\\

    
     BML \cite{zhou2021binocular} &  ICCV’21   & 67.04±0.63 & 83.63±0.29 &68.99±0.50 &85.49±0.34 & SetFeat\cite{afrasiyabi2022matching} &  CVPR’22   &79.60±0.80  & 90.48±0.44 \\
  
    DeepBDC\cite{xie2022joint} &  CVPR’22   & 67.34±0.43& \textcolor{blue!90!white}{84.46±0.28} &72.34±0.49&87.31±0.32  & 
    MetaNODE \cite{zhang2022metanode}  &  AAAI’22   & 80.82±0.75 &91.77±0.49\\
    
     MELR \cite{fei2020melr} &  ICLR’21  & 67.40±0.43& 83.40±0.28 &72.14±0.51 &87.01±0.35 & FRN* \cite{wertheimer2021few} &  CVPR’21 & 82.55±0.19&92.98±0.10 \\

    SetFeat\cite{afrasiyabi2022matching} &  CVPR’22   & 68.32±0.62& 82.71±0.46 &\textcolor{blue!90!white}{73.63±0.88}&\textcolor{blue!90!white}{87.59±0.57} & 
    RENet \cite{dong2023exploring} &  AAAI’23 & 83.33±0.40&92.97±0.24 \\

    ESPT\cite{rong2023espt} &  AAAI’23   & \textcolor{blue!90!white}{68.36±0.19}& 84.11±0.12 &72.68±0.22&87.49±0.14 & 
    DeepBDC\cite{xie2022joint} &  CVPR’22   &\textcolor{blue!90!white}{83.55±0.40}  & \textcolor{blue!90!white}{93.82±0.17} \bigstrut[b]\\
    
    \hline
    \rowcolor{gray!20} MLCC  & Ours $\uparrow$   & \textbf{69.04±0.46} &\textbf{85.61±0.26} & \textbf{74.05±0.49} & \textbf{89.02±0.29}  & MLCC  & Ours $\uparrow$  & \textbf{84.21±0.39} &\textbf{94.34±0.17} \bigstrut[t]\\
      
    \bottomrule
  \end{tabular}
\vspace{-0.1in}
\end{table*}

\begin{table}[h]
  \small
  \centering
  \renewcommand\tabcolsep{3pt}
  \renewcommand\arraystretch{1.1}
  \caption{Comparison with state-of-the-art methods for 5-way 5-shot classification in cross-domain scenarios.}
  \vspace{-0.1in}
  \label{Tab_2}
  \vspace{0.01in}
  \begin{tabular}{l|c|c|c|c}
  \toprule
    
    \multicolumn{1}{l|}{\multirow{2}{*}{Methods}} 
    & \multicolumn{1}{c|}{\multirow{2}{*}{ResNet-12}} 
    & \multicolumn{1}{c|}{miniImage.} 
    & \multicolumn{1}{c|}{miniImage.}
    & \multicolumn{1}{c}{miniImage.}\\
    \multicolumn{1}{l|}{}  &  &$\rightarrow$ Car  & $\rightarrow$ Aircraft  & $\rightarrow$ CUB \bigstrut[b]\\
    
    \hline
    ProtoNet \cite{snell2017prototypical}  &NIPS’17  & 46.30±0.36 & 55.96±0.38 & 67.19±0.38 \bigstrut[t] \\
    GEmbed \cite{tian2020rethinking} &  ECCV’20  & 50.18±0.37 & 58.95±0.38 & 67.43±0.44 \\
    CovNet \cite{wertheimer2019few}  &  CVPR’19  & 52.90±0.37  & 63.56±0.37 & 76.77±0.34 \\
    ADM \cite{li2020asymmetric} & IJCAI’20 & 53.94±0.35 & 65.40±0.36 & 70.55±0.43 \\
    DeepBDC\cite{xie2022joint} &  CVPR’22 & \textcolor{blue!90!white}{54.61±0.37}  & \textcolor{blue!90!white}{68.67±0.39} & \textcolor{blue!90!white}{77.87±0.33}  \bigstrut[b]\\
    \hline
    \rowcolor{gray!20} MLCC  & Ours $\uparrow$ & \textbf{56.43±0.37} & \textbf{69.75±0.42} & \textbf{80.44±0.35} \bigstrut[t]\\
      
    \bottomrule
  \end{tabular}
\end{table}

\subsection{Implementation Details}
We adopt the same architecture used in some recent work \cite{xie2022joint}, which aims to enable fair comparisons with previous methods. Specifically, two popular network architectures ResNet-12
and ResNet-18 are selected as backbones, with input image sizes of $84 \times 84$ and $224  \times  224$, respectively. 
We initialize the model weights by conducting a pre-training stage.
In each meta-training/testing episode task, we compute euclidean distance between the prototypes and query samples to make classification decisions for each query sample. Our code is publicly available at 
\href{https://github.com/Sakura65/ICME2024_MLCC}{https://github.com/ourlab/icme24mlcc}.


\subsection{Comparisons with State-of-The-Art}
To verify the effectiveness of the proposed MLCC method, we conduct three types of experiments to validate model performances compared with existing state-of-the-art baselines. All comparative results are summarized in Table~\ref{Tab_1} and Table~\ref{Tab_2}, we can observe that 
1) \textbf{General few-shot image classification:} Our MLCC approach clearly outperforms all the comparative baselines on benchmark datasets, i.e., miniImageNet and tieredImageNet. Particularly, compared to the current best-performing methods in 5-way 5-shot metric, such as DeepBDC and SetFeat, MLCC achieves average improvements of 1.15\% and 1.43\%.
2) \textbf{Fine-grained few-shot image categorization:} Meanwhile, our proposed approach achieved the highest performance on the CUB dataset, reaching accuracy rates of 84.21\% and 94.34\% in 1-shot and 5-shot settings, respectively. This phenomenon demonstrates that our method can excavate more comprehensive and accurate semantic concepts in fine-grained scenarios. 
3) \textbf{Cross-domain few-shot image recognition:} Furthermore, we extend our experimental evaluation in more challenging cross-domain scenarios. The results show that our method can be more easily extended to domain shift scenarios than the previous baselines.
In general, these experimental results provide compelling evidence regarding the efficacy and robustness of our MLCC method across a range of visual recognition experimental settings.

\begin{table}[t]
\small
\centering
\renewcommand\tabcolsep{3.7pt}
\renewcommand\arraystretch{1.1}
\caption{Ablation studies for our MLCC on miniImageNet.}
\vspace{-0.1in}
\label{Tab_3}
\begin{tabular}{c|ccc|cc}
         \toprule
          Settings & $\mathcal{L}_{\text{Intra-class}}$ & $\mathcal{L}_{\text{Inter-class}}$ 
          & $\mathcal{L}_{\text{Forget}}$
          & 5way-1shot & 5way-5shot \bigstrut[b] \\
          \hline
          I &\cmark  &\cmark  &\xmark   & 68.32±0.45 & 84.43±0.28  \bigstrut[t]\\
          II &\xmark &\cmark  &\cmark     & 67.63±0.44 &84.72±0.27  \bigstrut[t]\\
          III &\cmark &\xmark   &\cmark  & 67.81±0.43  &84.83±0.27 \bigstrut[b]\\
          IV &\xmark &\xmark  &\cmark  & 68.12±0.44  & 85.07±0.28 \bigstrut[b]\\
          \hline
          \rowcolor{gray!20}  MLCC & \cmark  &\cmark  &\cmark & \textbf{69.04±0.46} &\textbf{85.61±0.26} \bigstrut[t]   \\
          \bottomrule
     \end{tabular}
\end{table}

\begin{figure}[t]
  \centering
  \begin{subfigure}{0.49\linewidth}
    \includegraphics[width=0.99\linewidth]{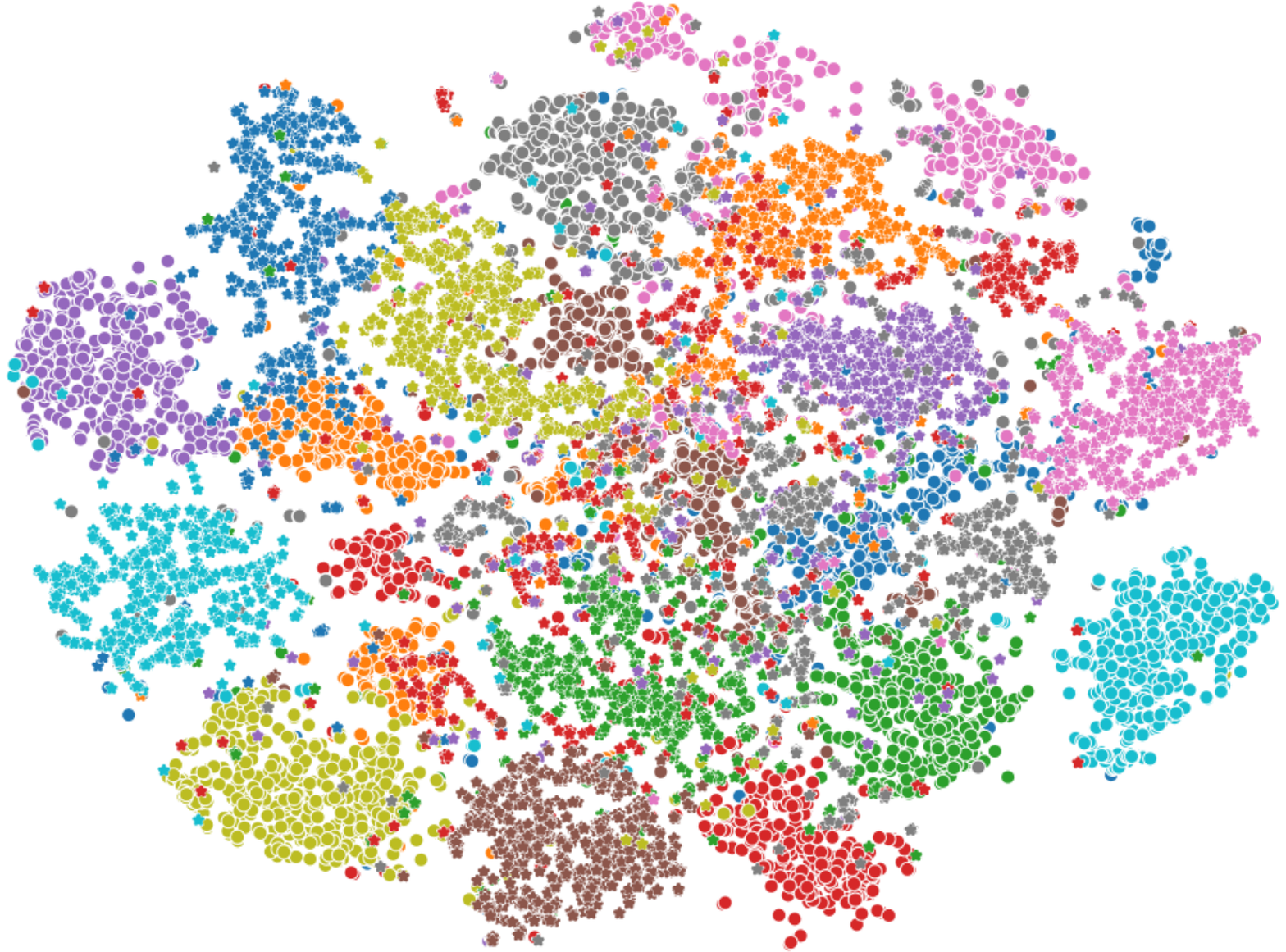}
    \caption{ResNet-12}
    
  \end{subfigure}
  \hfill
  \begin{subfigure}{0.49\linewidth}
    \includegraphics[width=0.99\linewidth]{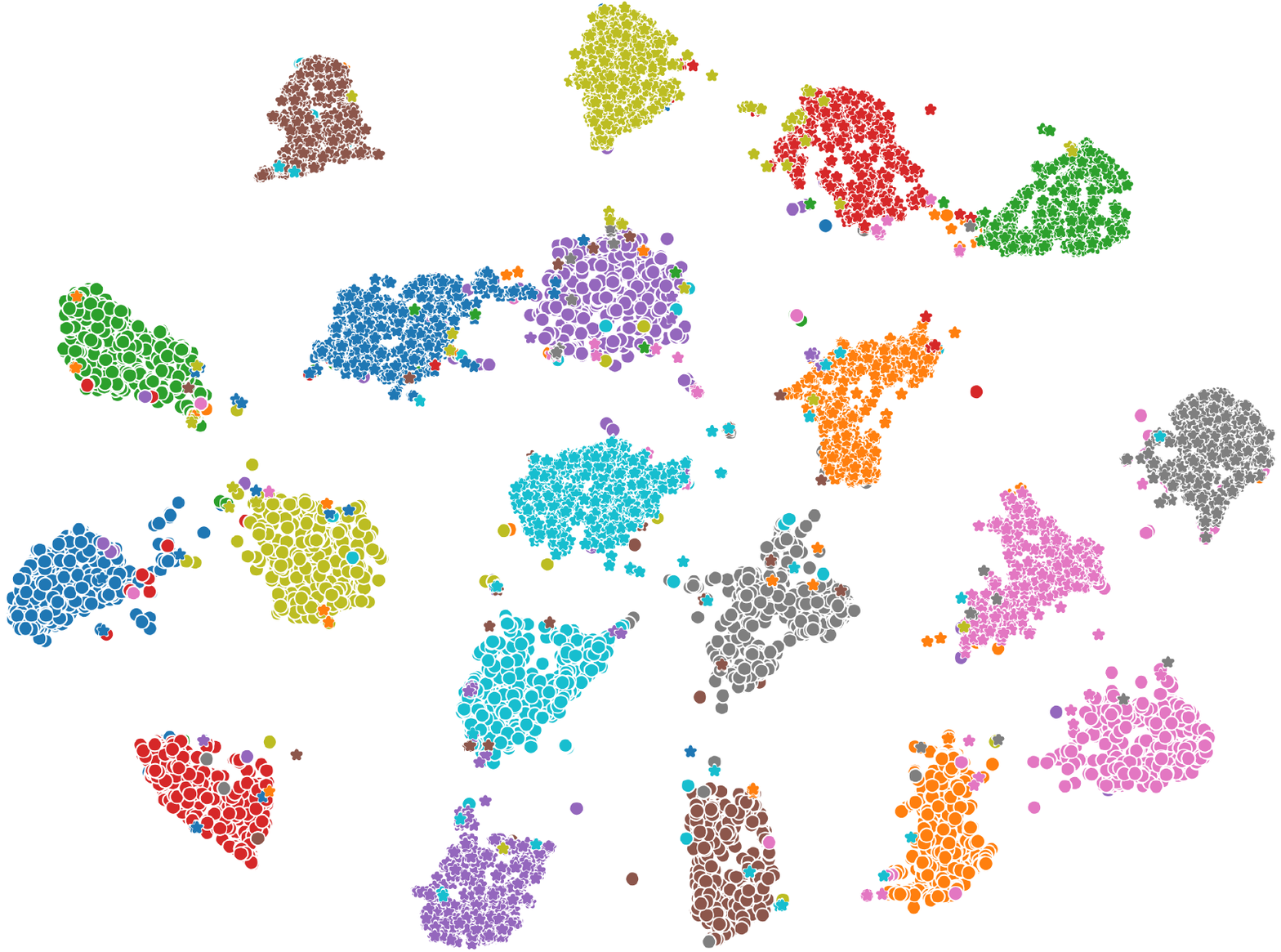}
    \caption{Our}
    
  \end{subfigure}
  \vspace{-1mm}
  \caption{T-SNE visualization of the discriminative features
learned by ResNet-12 and our approach for novel classes.}
  
  \vspace{-2mm}
  \label{Fig_3}
\end{figure}
\subsection{Ablation Studies}
To validate and further investigate the benefits of the proposed method, we conduct ablation studies on different module combinations derived from MLCC. The comparative results are summarized in Table~\ref{Tab_3}. The results of ``MLCC-I" clearly indicate that excluding the Forget module leads to reduced robustness,  highlighting the crucial role of the Forget module in exploring the previous prior knowledge. When we exclude the intra- and inter-class constraints, or both from MLCC separately, we observe an overall decrease in the model's performance, which is correspondingly denoted by ``MLCC-II", ``MLCC-III", and ``MLCC-IV". The above results of the ablation studies demonstrate the importance of each component in our MLCC and their collective integration to combat inductive bias and catastrophic forgetting.

\subsection{Visualization Results}
We utilize visualization techniques to effectively demonstrate the superior capacity of our MLCC approach in capturing robust and meaningful features. Concretely, the T-SNE is used to portray the distribution of feature representations extracted from novel classes. As depicted in Fig.~\ref{Fig_3}, the T-SNE  resulting from our MLCC method exhibits an evident increase in intra-class compactness, while revealing a notable improvement in the separation between different classes.

\section{Conclusion}
\label{sec:conclusion}
Our proposed MLCC approach offers a promising solution in few-shot learning by tackling the challenges of inductive bias and catastrophic forgetting. Through a novel multi-level contrastive constraints framework, we remove the error accumulation caused by inductive bias and alleviate the catastrophic forgetting of priors, leading to the correct inductive paradigm for each class and enhancing the learning of previous prior knowledge. The substantial boosts consistently demonstrate the efficiency of MLCC for few-shot image classification.

\vspace{1.3em}
\textbf{Acknowledgment:}
This work was supported in part by the National Natural Science Foundation of China (Grant Nos. 62302172, 62176077), in part by the Guangdong University Young Innovative Talents Program Project (Grant No. 2023KQNCX020), in part by the Guangdong International Science and Technology Cooperation Project (NO. 2023A0505050108), and in part by the Guangdong International Science and Technology Cooperation Project (Grant No.2023A0505050108).


\small
\bibliographystyle{IEEEbib}
\bibliography{icme2023template}

\end{document}